\documentclass[aps,prl,showpacs,groupedaddress,twocolumn,preprintnumbers,amsmath,amssymb]{revtex4}

\usepackage{graphicx}% Include figure files
\usepackage{dcolumn}% Align table columns on decimal point
\usepackage{bm}% bold math
\usepackage{braket}
\usepackage{color}

\begin{document}

\title{Spinor dynamics in a mixture of spin-1 and spin-2 Bose-Einstein condensates}
% Force line breaks with \\

\author{Yujiro Eto$^{1}$}
\author{Hitoshi Shibayama$^{2}$}
\author{Hiroki Saito$^{3}$}
\author{Takuya Hirano$^{2}$}
\affiliation{%
$^{1}$National Institute of Advanced Industrial Science and Technology (AIST), NMIJ, Tsukuba 305-8568, Japan\\
$^{2}$Department of Physics, Gakushuin University, Toshima, Tokyo 171-8588, Japan\\
$^{3}$Department of Engineering Science, University of Electro-Communications, Chofu, Tokyo 182-8585, Japan}

\date{\today}% It is always \today, today,
             
\begin{abstract}
The spinor dynamics of Bose-Einstein condensates of $^{87}$Rb atoms with hyperfine spins 1 and 2 were investigated.
A technique of simultaneous Ramsey interferometry was developed for individual control of the vectors of two spins with almost the same Zeeman splittings.
The mixture of spinor condensates is generated in the transversely polarized spin-1 and the longitudinally polarized spin-2 states.
The time evolution of the spin-1 condensate exhibits dephasing and rephasing of the Larmor precession due to the interaction with the spin-2 condensate.
The scattering lengths between spin-1 and -2 atoms were estimated by comparison with the numerical simulation using the Gross-Pitaevskii equation.
The proposed technique is expected to facilitate the further study of multiple spinor condensates.
\end{abstract}

\maketitle
Atomic Bose-Einstein condensates (BECs) with internal spin degrees of freedom, so-called spinor BECs, have attracted much attention as intriguing quantum spin systems that have various magnetic and dynamic properties due to spin-dependent atomic interactions \cite{Kawaguchi12,Kurn13}.
The magnetic phase of the ground state of spin-1 BEC can be ferromagnetic or antiferromagnetic \cite{Ohmi98,Lo98}.
Even richer phase diagrams are predicted for larger spins, such as the cyclic phase in a spin-2 BEC \cite{Koashi00,Ciobanu00,Klausen01,Ueda02}.
A variety of spinor dynamics related to ferromagnetism or antiferromagnetism have been experimentally reported \cite{Schmaljohann04,Chang04,Kuwamoto04,Sadler06,Black07,Liu09,Kronjager10,Vint13,Fang16,Anquez16,Seo16,Naylor16}.
Magnetic dipolar interactions, which are anisotropic long-range interactions, and quantum spin noise further enrich the spinor properties of BECs.
Spinor dipolar effects, such as spin texture formation \cite{Vengalattore08,Eto14PRL}, spontaneous demagnetization \cite{Pasquiou11,Pasquiou12,Naylor15}, magnon energy gap \cite{Marti14}, and quantum spin squeezing and entanglement \cite{Hamley12,Linnemann16} have been observed.

A wide variety of spinor dynamics have been investigated using spin-1, -2 or -3 BECs.
However, experiments on mixtures of multiple spins have not been reported to date.
Novel ground states and spinor dynamics can be expected due to the interactions between multiple spins.
Theoretically, the existence of a ground state with a broken-axisymmetry phase has been predicted in a mixture of two atomic spin-1 BECs \cite{Xu10}.

\begin{figure}[tbp]
\includegraphics[width=8cm]{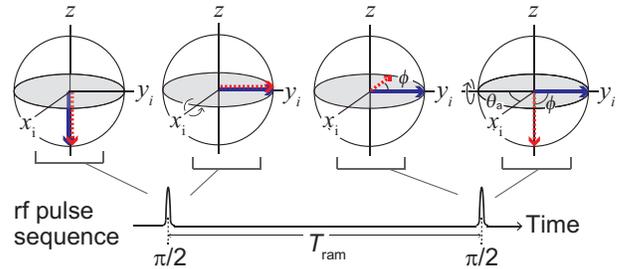}
\caption{
(color online) Schematic illustrations of simultaneous Ramsey interferometry to control two hyperfine spins of $^{87}$Rb atoms.
The solid and dotted arrows indicate the directions of the spin-1 and spin-2 vectors, respectively, and $x_i$-$y_i$ is the rotating reference frame for spin $i$.
The first $\pi/2$ pulse rotates spin $i$ from the $-z$ direction to the $y_i$ direction.
After the angle between the two spin vectors becomes $\phi$ due to the different Larmor frequencies, the second $\pi/2$ pulse is applied.
The angle between the two $\pi/2$ rotation axes is $\theta_a$. ($\phi = \theta_a = \pi / 2$ in the illustration.)
}
\end{figure}

In this Letter, we investigate the spinor dynamics in a mixture of spin-1 and -2 BECs using the two controlled hyperfine spins of $^{87}$Rb atoms.
The Zeeman splittings of the two hyperfine spins are almost the same in the weak magnetic field; therefore, it is difficult to control each spin individually in a conventional manner, such as Rabi rotation.
Here we propose a novel scheme to rotate two spins in arbitrary directions, which uses the simultaneous Ramsey-type interferometry of two hyperfine spins.
Using this technique, a spinor mixture of the transversely polarized spin-1 and the longitudinally polarized spin-2 states is generated, and the dephasing and rephasing of the Larmor precession in a spin-1 BEC are observed.
We show that this phenomenon is induced by the interaction between spin-1 and -2 BECs.
In addition, the scattering lengths between spin 1 and 2 of $^{87}$Rb atoms are determined by comparison with the numerical simulation using the Gross-Pitaevskii (GP) equation.

\begin{figure}
\includegraphics[width=8cm]{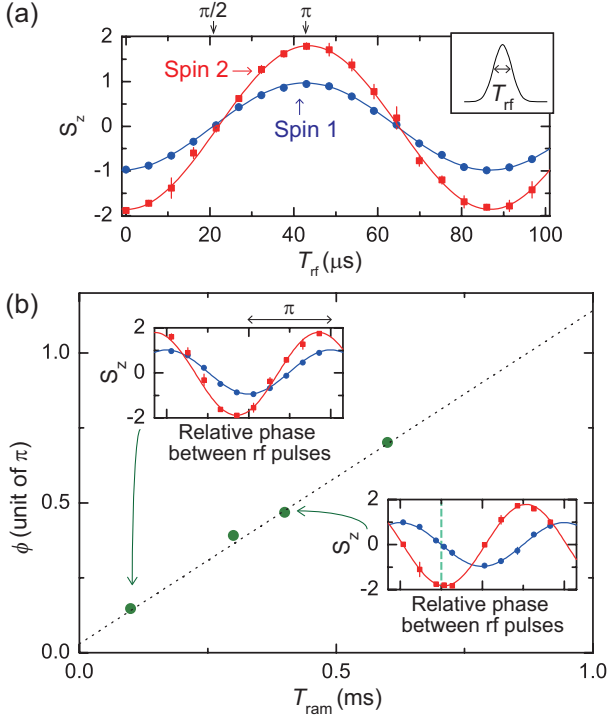}
\caption{
(color online) Experimental control of spin-1 and -2 BECs.
(a) Simultaneous rotation of spins 1 and 2 by a single rf pulse. 
$S_{z}$ is measured as $T_{\mathrm {rf}}$ is changed.
The inset shows that the envelope of rf pulses is Gaussian, and $T_{\mathrm {rf}}$ corresponds to twice the standard deviation. 
(b) Control of spins using a simultaneous Ramsey interferometer.
The relative angle between two spins $\phi$, is plotted as a function of $T_{\mathrm {ram}}$.
The slope of the dotted line is $\Delta_\Omega$.  
The insets show $S_z$ measured using the Ramsey interferometer as a function of the relative phase between rf pulses at fixed $T_{\mathrm {ram}}$ of $0.1$ and $0.4$ ms.
The vertical dashed line in the inset corresponds to the rightmost state in Fig. 1.
In (a) and in the insets of (b), each plotted point and error bar represent an average and standard deviation over three measurements.
}
\end{figure}  
\begin{figure}
\includegraphics[width=8cm]{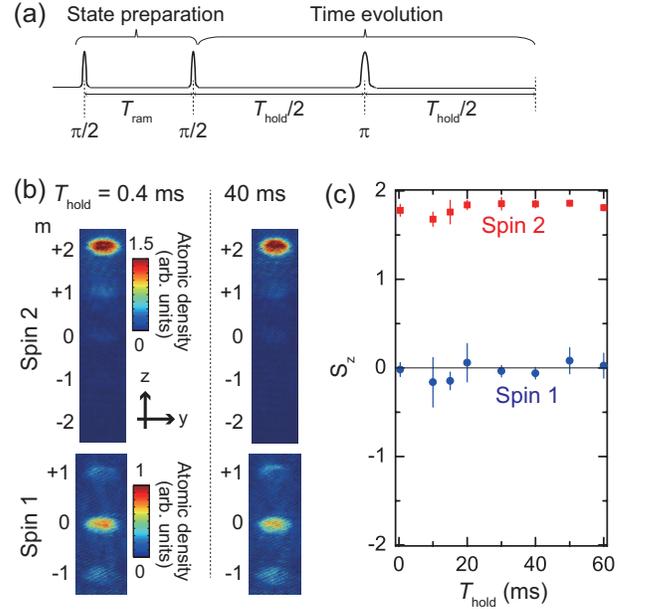}
\caption{
(color online) Time evolution of the spinor mixture of the transversely polarized spin-1 and the longitudinally polarized spin-2 states.
(a) Rf pulse sequence with $T_{\mathrm{ram}} =$ 0.4 ms.
The $\pi$ pulse is applied to cancel out the effect of inhomogeneous magnetic field.
(b) Absorption images of each hyperfine sublevel $m$ taken at $T_{\mathrm{hold}} =$ 0.4 and 40 ms.
(c) $S_z$ as a function of $T_{\mathrm{hold}}$.
}
\end{figure}

We briefly explain the scheme to simultaneously control the two hyperfine spins.
The Larmor frequencies in spin-1 and -2 atoms are calculated to be $\Omega_{1}/(2\pi) \sim 140.1$ kHz and $\Omega_{2}/(2\pi) \sim139.5$ kHz, respectively, in a bias magnetic field of $ B_z = 199.5$ mG.
The Larmor precessions therefore change the relative azimuthal angle between spins as $(\Omega_1 - \Omega_2) t \equiv \Delta_\Omega t$.
By combining these simultaneous Larmor precessions and the spin rotation using the resonant radio frequency (rf) pulses,
the spin vectors can be turned in arbitrary directions.
A specific procedure to generate the spinor mixture in the transversely polarized spin-1 and the longitudinally polarized spin-2 states is depicted in Fig. 1.
Let us consider the situation in which both spin-1 and -2 atoms are oriented in the $-z$ direction, where the bias magnetic field, $B_z$, is applied in the $z$ direction.
The first $\pi/2$ pulse with frequency $\Omega_{1}$ rotates the spin 1 around the axis $\bm{e}_x \cos{\Omega_{1} t} - \bm{e}_y \sin{\Omega_{1} t}$ and spin 2 around an axis $- \bm{e}_x \cos{\Omega_{1} t} -  \bm{e}_y \sin{\Omega_{1} t}$ because the $g$ factors for the spin-1 and -2 atoms have opposite signs.
Different reference frames of $x_{1}$-$y_{1}$ and $x_2$-$y_2$ are adopted for spin 1 and spin 2 vectors that rotate in the opposite directions at frequency $\Omega_1$.
The second $\pi/2$ pulse rotates the spin-$i$ vector around the $\bm{e}_{x_i} \cos{\theta_a} - \bm{e}_{y_i} \sin{\theta_a}$ axis,
where $i = 1$ or $2$ and $\theta_a$ can be controlled by the relative phase between the first and second $\pi$/2 pulses.
The transversely polarized spin-1 and the longitudinally polarized spin-2 states are generated when the time separation between the first and second $\pi$/2 pulses, $T_{\mathrm{ram}}$, is set to approximately $\pi / (2 \Delta_\Omega)$ and $\theta_a = \pi/2$, as illustrated in Fig. 1.
By controlling the intensity, timing, and the relative phase between the two rf pulses, spins in arbitrary directions can be generated.

An $^{87}$Rb BEC containing $3.1(2)\times10^5$ atoms in the hyperfine state $\ket{f = 2, m = -2}$ is produced in a crossed far-off-resonant optical dipole trap with axial and radial frequencies of $\omega_z / (2\pi) = 30$ Hz and $\omega_r / (2\pi) =174$ Hz (see Ref. \cite{Eto13PRA} for a more detailed description).
The external magnetic field of $B_z = 199.5$ mG is aligned with the axis of the trap ($z$ direction).
A mixture of $\ket{2, -2}$ and $\ket{1, -1}$ with a ratio of approximately $6:4$ is generated by application of the resonant microwave pulse, which corresponds to the leftmost panel in Fig. 1.
The rf pulse sequence is applied and the BECs are then released from the trap.
After the Stern-Gerlach separation,
the images of spin-2 and -1 atoms are taken with a time-of-flight expansion for 13 ms and 15 ms, respectively \cite{Tojo10}.
The expectation value of the spin-$z$ component is obtained as $S_{z} = \Sigma_{m} m N_{f, m}/\Sigma_{m} N_{f, m}$ with $f = 1$ or 2,
where $N_{m}$ is the atomic number in the hyperfine state $\ket{f, m}$.

\begin{figure*}[t]
\includegraphics[width=16.5cm]{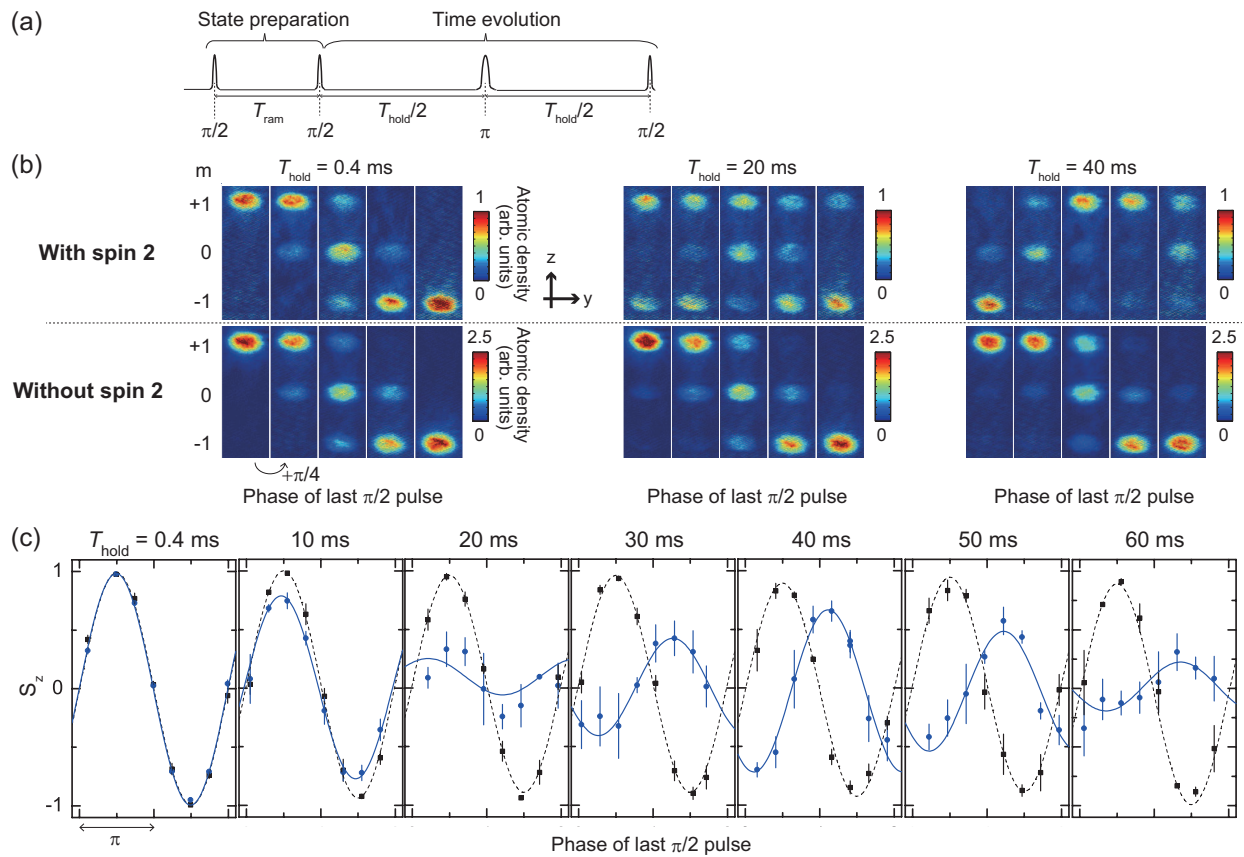}
\caption{
Effect of spin-2 BEC on the spin-1 dynamics.
(a) Rf pulse sequence. 
A spinor mixture is prepared in the transversely polarized spin-1 and the longitudinally polarized spin-2 states, as shown in Fig. 3.
The final $\pi/2$ pulse is applied to convert $S_x$-$S_y$ to $S_z$.
(b) Absorption images of the spin-1 BEC with and without the spin-2 BEC for $T_{\mathrm{hold}} =$ 0.4, 20, and 40 ms.
(c) $S_z$ values for the spin-1 BEC as a function of the phase of the final $\pi$/2 pulse. 
The circles and squares (fitted by solid and dashed sinusoidal curves) indicate the results with and without the spin-2 BEC, respectively.
$T_{\mathrm{hold}}$ is set to 0.4 10, 20, 30, 40, 50, and 60 ms. 
}
\end{figure*}

We first confirmed that the Rabi oscillations occur simultaneously for spins 1 and 2, as shown in Fig. 2(a).
A single rf pulse is applied to the mixture of $\ket{2, -2}$ and $\ket{1, -1}$ states while changing the pulse duration, $T_{\mathrm {rf}}$.
The measured $S_{z}$ for the spin-1 and -2 BECs oscillate with an equal frequency.
Irradiation of rf pulses with $T_{\mathrm {rf}} = 21.5$ and $43.0$ $\mu$s was determined to correspond to the $\pi$/2 and $\pi$ pulses, respectively.

The spin-1 and spin-2 vectors were experimentally controlled using the Ramsey sequence in Fig. 1.
The insets in Fig. 2(b) show $S_z$ measured using the Ramsey interferometer as a function of the relative phase between the rf pulses.
The relative angle of spins $\phi$, can be obtained from the phase shift between the sinusoidal curves in the inset.
The observed values of $\phi$ are proportional to $T_{\mathrm {ram}}$, as shown in Fig. 2(b), and the slope is in agreement with $\Delta_\Omega$ (the dotted line in Fig. 2(b)).
The two spins become almost orthogonal [$\phi = (0.468\pm0.005) \times \pi$], when $T_{\mathrm {ram}} = 0.4$ ms.
By setting the relative phase between the rf pulses to the dashed line in the right-hand inset of Fig. 2(b), the transversely polarized spin-1 and longitudinally polarized spin-2 states are realized.

\begin{figure}[t]
\includegraphics[width=8cm]{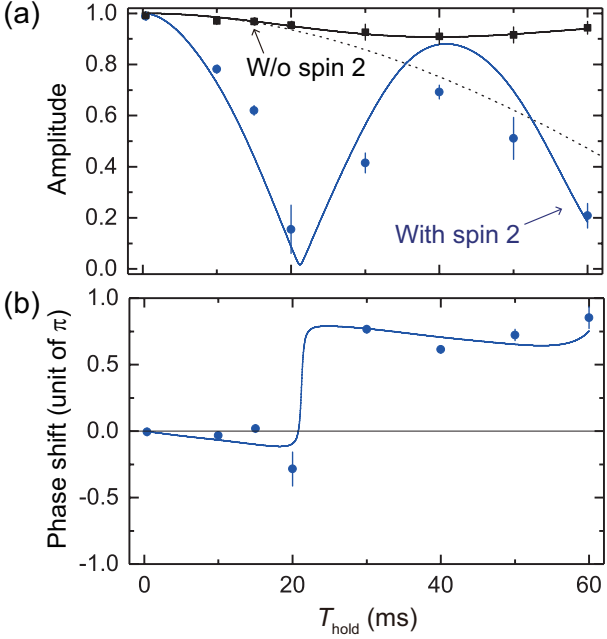}
\caption{
(color online) Effects of the spin-2 BEC on the dynamics of the spin-1 BEC.
(a) Larmor-precession amplitude of the spin-1 BEC.
The circles and squares indicate the experimental values with and without spin-2 BECs, respectively.
The solid curves are obtained by numerical simulation of the GP equation.
The dotted curve in (a) indicates the amplitude decay only due to the quadratic Zeeman effect. 
(b) Shift in the Larmor-precession angle of the spin-1 BEC due to interaction with the spin-2 BEC.
The experimental values in (a) and (b) are obtained by sinusoidal fitting in Fig. 4(c).
}
\end{figure}

The time evolution of the $\ket{f, m}$ populations was investigated by holding the spinor mixture of the transversely polarized spin-1 and the longitudinally polarized spin-2 states during $T_{\mathrm {hold}}$.
The $\pi$ pulse applied at the middle of the holding time, as shown in Fig. 3(a), induces spin echo, and suppresses the effects of the spatial inhomogeneity of the magnetic field, e.g., the formation of helical spin and $m$-dependent position shift \cite{Eto13APEX,Eto14}.
As shown in Figs. 3(b) and 3(c), the populations and $S_z$ are almost independent of $T_{\mathrm{hold}}$.
This indicates that the spin exchanging elastic collisions between spin 1 and 2, e.g., $(\ket{1,0}, \ket{2,-2}) \rightarrow (\ket{1,-1}, \ket{2,-1})$, hardly occur, 
because such processes are incompatible with the conservation of the Zeeman energy due to the different $g$ factors for the spin-1 and -2 atoms.

The dynamics of the spin-1 state in the $x_1$-$y_1$ plane were investigated using the sequence in Fig. 4(a). 
The spin evolution in the $x_1$-$y_1$ plane is sensitive to the temporal fluctuation of the magnetic field,
and the spin-echo $\pi$ pulse also suppresses the effect of run-to-run fluctuation of the magnetic field in addition to the effects of spatial inhomogeneity \cite{Eto13APEX}.
The $\pi/2$ pulse at the end of the sequence rotates the spin $i$ around the $\bm{e}_{x_i} \cos{(\theta_b+\pi/2)} - \bm{e}_{y_i} \sin{(\theta_b+\pi/2)}$ axis,
where $\theta_b$ can be controlled by the phase of the $\pi/2$ pulse.
Thus, by measuring $S_z$ as a function of $\theta_b$, as shown in Figs. 4(b) and 4(c), 
the precession amplitudes and the azimuthal angle of the spin vector in the $x_1$-$y_1$ plane just before the final $\pi/2$ pulse can be obtained.

First we discuss the dynamics of the spin-1 BEC without the spin-2 BEC.
The plots in Fig. 5(a) show the precession amplitudes of the spin-1 BEC obtained from the sinusoidal fitting in Fig. 4(c).
The precession amplitude of the BEC remains close to unity, even at 60 ms.
The long spin precession of the spin-1 $^{87}$Rb BEC has already been observed, as reported in Ref. \cite{Higbie05}.
If there were no atomic interactions in the spin-1 BEC, then the phase shift due to the quadratic Zeeman effect would cause a larger amplitude decay, as shown by the dotted curve in Fig. 5(a). The suppression of the amplitude decay is therefore due to the ferromagnetic nature of spin-1 $^{87}$Rb.

The effects of the interaction between the spin-1 and -2 BECs are clearly shown in Fig. 5.
In the spinor mixture system, the precession amplitude of spin 1 is modulated and the azimuthal angles are shifted due to the interaction with spin-2 BECs, as shown in Figs. 5(a) and 5(b).
At $T_{\mathrm {rf}} = 20$ ms, the amplitude is maximally suppressed, where the $m = \pm1$ states are equally populated with small $m = 0$ population, as shown in Fig. 4(b). 
Such population distributions cannot be produced only by the spin rotation of the polarized spin-1 state by the rf pulse sequence. 

To understand the observed spinor dynamics, we consider the mean-field evolution of the system described by the macroscopic wave functions $\psi_{fm}(\bm{r})$, of the hyperfine state, $\ket{f, m}$.
The one-body part of the mean-field energy is given by $E_1 = \int d\bm{r} \sum_{f, m} [\frac{\hbar^2}{2M} |\nabla\psi_{fm}|^2 + (V + g_f \mu_B m B_z
+ q_f m^2 B_z^2) |\psi_{fm}|^2]$, where $M$ is the mass of the $^{87}{\rm Rb}$ atom, $V$ is the trap potential, $\mu_B$ is the Bohr magneton, and $g_f$ and $q_f$ are the $g$-factor and quadratic Zeeman coefficient of the spin-$f$ state, respectively.
The $s$-wave interaction energy is expressed as $E_s = \int d\bm{r} \sum_{f, f'} \frac{2\pi\hbar^2 a_{\cal F}^{(f, f')}}{M} \sum_{{\cal F, M}} |\sum_{m, m'} \braket{{\cal F, M} | f, m; f', m'}$ $\psi_{fm} \psi_{f'm'}|^2$, where $a_{\cal F}^{(f, f')}$ is the $s$-wave
scattering length between atoms with spin $f$ and $f'$ for colliding channels with total spin ${\cal F}$, and $\braket{{\cal F, M} | f, m; f', m'}$ is the Clebsch-Gordan coefficient.
We also take into account the magnetic dipole-dipole interaction energy, $E_{d} = \int d\bm{r} d\bm{r}' \sum_{f, f'}
\frac{\mu_0 \mu_B^2 g_f g_{f'}}{8\pi |\bm{r} - \bm{r'}|^3} \{ \bm{F}^{(f)}(\bm{r}) \cdot \bm{F}^{(f')}(\bm{r}') - 3 [\bm{F}^{(f)}(\bm{r}) \cdot \bm{e}] [\bm{F}^{(f')}(\bm{r}') \cdot \bm{e}] \}$, where $\mu_0$ is the magnetic permeability of a vacuum, and $\bm{F}^{(f)} = \sum_{m, m'} \psi_{fm}^* \bm{S}_{mm'}^{(f)} \psi_{fm'}$ where $\bm{S}^{(f)}$ is the vector of the
$(2f + 1) \times (2f + 1)$ spin matrices.
The GP equation, $i \hbar \partial_t \psi_{fm} = \delta(E_1 + E_s + E_d) / \delta \psi_{fm}^*$, is numerically solved using the pseudospectral method.
The intra-spin scattering lengths are taken to be $a_2^{(1, 1)} - a_0^{(1, 1)} = -1.4 a_B$, $a_2^{(2, 2)} - a_0^{(2, 2)} = 3.35 a_B$, and $a_4^{(2, 2)} - a_2^{(2, 2)} = 7.7 a_B$~\cite{Kempen}, where $a_B$ is the Bohr radius.
The inter-spin scattering lengths are determined to be $a_3^{(1, 2)} - a_2^{(1, 2)} = 1.8 a_B$ and $a_1^{(1, 2)} - a_2^{(1, 2)} = 2.3 a_B$ in such a way that the numerical curves best fit the experimental data in Fig. 5.

The effect of interaction between spins 1 and 2 observed in Fig. 5 can be understood as follows. 
If we assume that the spin-2 atoms are fixed to the $m = -2$ state ($m = 2$ after the $\pi$ pulse), then the inter-spin interaction part of the Larmor-averaged GP equation reduces to $i \hbar \partial_t \psi_{1, m} \propto \frac{4\pi\hbar^2}{M} c_m
|\psi_{2, -2}|^2 \psi_{1, m}$, where $c_1 = (9 a_1^{(1, 2)} + 5 a_2^{(1, 2)} + a_3^{(1, 2)}) / 15$, $c_2 = (2 a_2^{(1, 2)} + a_3^{(1, 2)}) / 3$, and $c_3 = a_3^{(1, 2)}$. 
The transverse magnetization thus evolves as $F_+^{(1)} = F_x^{(1)} + i F_y^{(1)} \propto e^{i \alpha t} \cos(\beta t)$, where $\alpha = \frac{4\pi\hbar^2}{M} |\psi_{2, -2}|^2 (9 a_1^{(1, 2)} - 5 a_2^{(1, 2)} - 4 a_3^{(1, 2)}) / 15$ and $\beta = \frac{4\pi\hbar^2}{M} |\psi_{2, -2}|^2 2 (a_2^{(1, 2)} - a_3^{(1, 2)}) / 3$.
The dipole interaction and quadratic Zeeman effect also contribute to the change in the transverse magnetization.

In conclusion, the dynamics in a mixture of spinor BECs were investigated.
A novel technique that employs simultaneous interferometry is proposed to individually control the two hyperfine spins.
This technique was used to observe the dephasing and rephasing dynamics of the Larmor precession induced by the interaction between spins 1 and 2.
The experimental results were compared with numerical calculations and the scattering lengths between spin-1 and -2 $^{87}$Rb atoms were determined.
The proposed experimental technique and observed results are expected to stimulate the study of multiple spinor BECs.

This work was supported by Grants-in-Aid for Scientific Research (C) (Nos. 17K05595, 17K05596, 16K05505, and 15K05233), and a Grant-in-Aid for Scientific Research on Innovative Areas (No. 25103007, ``Fluctuation \& Structure") from the Ministry of Education, Culture, Sports, Science and Technology (MEXT), Japan.

\end{document}